# Simple Method for Measuring the Zero-Dispersion Wavelength in Optical Fibers

Maxime Droques, Benoit Barviau, Alexandre Kudlinski, Géraud Bouwmans and Arnaud Mussot

*Abstract*— We propose a very simple method for measuring the zero-dispersion wavelength of an optical fiber as well as the ratio between the third- and fourth-order dispersion terms. The method is based on the four wave mixing process when pumping the fiber in the normal dispersion region, and only requires the measurement of two spectra, provided that a source tunable near the zero-dispersion wavelength is available. We provide an experimental demonstration of the method in a photonic crystal fiber and we show that the measured zero-dispersion wavelength is in good agreement with a low-coherence interferometry measurement.

## I. INTRODUCTION

Group velocity dispersion (GVD) is one of the key characteristics of optical fibers. It is thus important to be able to accurately measure this parameter. The techniques developed to reach this goal can be divided into two main categories: the ones based on linear processes, such as time-of-flight, phase-shift or interferometric measurements [1-4]; and the ones based on nonlinear effects, such as four wave mixing (FWM), mainly [5-8]. The main advantage of these last ones is that the GVD measurement can be made in fiber samples ranging from a few meters up to hundred of meters long, while linear techniques are restricted to either very short samples (in the meter range) or to very long ones (in the kilometer range). In the intermediate length range, nonlinear methods provide accurate GVD measurements [5-8], but they usually require an important number of data, and especially an accurate knowledge of the fiber nonlinear coefficient and of the launched pump peak power.

In this work, we propose an extremely simple nonlinear method that requires the measurement of only two spectra to retrieve the zero-dispersion wavelength (ZDW, also labeled $\lambda_0$ in the text) of an optical fiber as well as the ratio between the third- and fourth-order dispersion terms. The main advantage of our method is the quasi-insensitivity of the FWM side lobes spectral position when the fiber is pumped in the normal GVD region. Moreover, we can retrieve back the whole GVD curve in the particular case of pure silica photonic crystal fibers (PCFs). Our measurement is found to be in good agreement with the ones from an usual low-coherence interferometric technique. The analytical model and assumptions are presented in section II and experimental results are shown in section III.

## II. MODEL

In the degenerate FWM process, energy transfers arising from the pump (located at $\omega_P$) to the signal and idler waves (respectively located at $\omega_S$ and $\omega_I$) are optimized when linear effects are perfectly balanced by nonlinear ones. For isotropic and single-mode fibers, near the ZDW, the phase-matching condition can be written as follows [9]:

$$\beta_2(\omega_P)[\omega_P - \omega_{S,I}]^2 + \frac{1}{12}\beta_4(\omega_P)[\omega_P - \omega_{S,I}]^4 + 2\gamma P = 0 \quad (1)$$

where $P$ is the pump peak power, $\beta_2$ and $\beta_4$ are the second- and fourth- order dispersion coefficients at the pump wavelength, $\gamma$ is the nonlinear coefficient defined as $\gamma = n_2\omega/(cA_{eff})$, with $A_{eff}$ the effective mode area and $n_2$ the fiber nonlinear refractive index. It is worth noting that the fourth-order dispersion term plays a major role in low normal GVD regimes. Indeed, in this case, it strongly affects the shape of the modulation instability (MI) spectrum and allows the generation of largely detuned FWM sidebands [10-14]. As an illustration of this, the spectral evolution of the FWM sidebands obtained by solving Eq.(1) is plotted in Fig.1-(a) for different pump powers and for a realistic PCF whose parameters are listed in Fig.1's caption. Figure 1 illustrates the important discrepancy between the shapes of FWM spectra in both GVD regimes. For normal GVD, two sidebands appear away from the pump, and we can note that increasing the pump power does not significantly modify the spectral position of the FWM sidebands. In the case of an anomalous GVD pumping, classical MI side lobes appear very close to the pump and, on the contrary to the previous case, the characteristics of the FWM spectrum strongly depends on pump power.

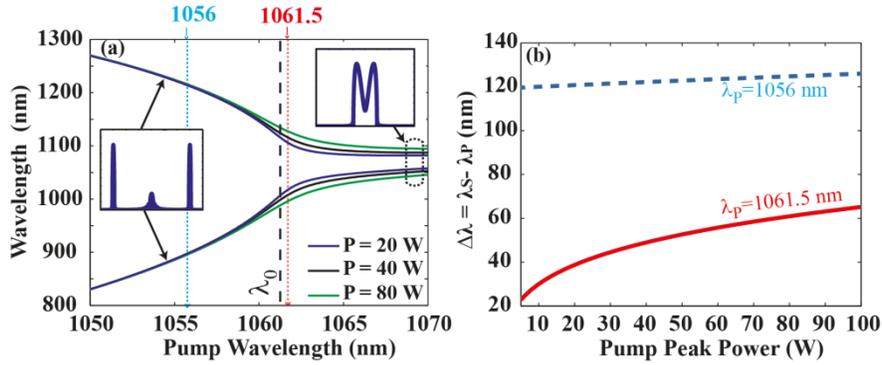

Figure 1: Wavelength of the FWM sidebands as a function (a) of pump wavelength and (b) of pump power for $\lambda_P$= 1061.5 nm (solid red curve) and $\lambda_P$= 1056 nm (dashed blue curve). The PCF structure is characterized by a pitch $\Lambda$ of 4.1 µm and a hole diameter $d$ of 2.6 µm. The GVD and nonlinear coefficient of the PCF have been calculated from ref [15]. $\lambda_0$ is the ZDW.

This is highlighted in Fig.1-(b), in which the detuning of the signal wavelength from the pump is plotted as a function of pump peak power for two pump wavelengths (1056 and 1061.5 nm) located on both sides of the ZDW (1061 nm). In the anomalous GVD regime, the side lobe detuning from the pump increases from 20 nm to 60 nm by increasing the pump power from 5 to 100 W. On the contrary, for a pump wavelength located in the normal GVD region, the spectral position of the signal wave is shifted by only 2 nm in the same power range. Our method is based on this particular property of the FWM process in the normal GVD regime since it allows the simplification of analytical calculations to retrieve the FWM position from Eq. (1). Thus, we firstly assume that the contribution of the nonlinear phase mismatch ($2\gamma P$) is negligible in Eq.(1). We will see in the following that this assumption is valid over an important range of pump powers. Secondly, since the pump wavelength is close to the ZDW, we assume that $\beta_4(\omega_p) \approx \beta_4(\omega_0)$, with $\omega_0 = 2\pi c/\lambda_0$. As can be seen from Fig.2-(a), this assumption leads to a discrepancy on the $\beta_4$ value of less than 5 % by pumping up to 10 nm below the ZDW. Finally, the third-order dispersion ($\beta_3$) is included by substituting $\beta_2(\omega_P)$ in Eq.(1) by its Taylor expansion around the zero-dispersion frequency:

$$\beta_2(\omega_P) = \beta_2(\omega_0) + \beta_3(\omega_0)[\omega_P - \omega_0] + \frac{\beta_4(\omega_0)}{2}[\omega_P - \omega_0]^2 \quad (2)$$

where $\beta_2(\omega_0)$ vanishes. We then obtain an equation with two unknown parameters, which are the ratio $\beta_3(\omega_0)/\beta_4(\omega_0)$ and $\lambda_0$, provided that the position of the FWM side bands as a function of the pump wavelength is known. The measurement of only two signal detunings ($\Delta\omega_{S1} = \omega_{P1} - \omega_{S1}$ and $\Delta\omega_{S2} = \omega_{P2} - \omega_{S2}$) corresponding to two pump frequencies ($\omega_{P1}$ and $\omega_{P2}$) thus allows to solve this system and leads to Eq. (3) and (4):

$$\frac{\beta_3(\omega_0)}{\beta_4(\omega_0)} = \frac{1}{12(\omega_{P1}-\omega_{P2})}\left\{\left[\left(\frac{\Delta\omega_{S1}+\Delta\omega_{S2}}{\omega_{P1}-\omega_{P2}}\right)^2+6\right]\left[\left(\frac{\Delta\omega_{S1}-\Delta\omega_{S2}}{\omega_{P1}-\omega_{P2}}\right)^2+6\right]\right\}^{1/2} \quad (3)$$

and

$$\lambda_0 = 2\pi c\left\{\frac{\beta_3(\omega_0)}{\beta_4(\omega_0)}+\frac{1}{12}(\omega_{P1}-\omega_{P2})^{-1}\left[6(\omega_{P1}^2-\omega_{P2}^2)+\Delta\omega_{S1}^2-\Delta\omega_{S2}^2\right]\right\}^{-1} \quad (4)$$

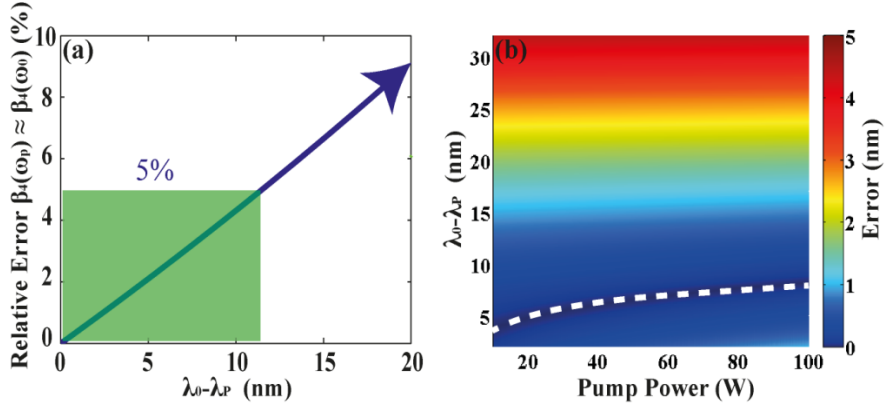

Figure 2: (a) Relative error between the fourth-order dispersion coefficient at the pump wavelength and at the ZDW as a function of the wavelength shift between the pump and the ZDW. (b) Error on the ZDW obtained by the present method as a function of pump power and detuning between the pump and the ZDW of a known GVD curve. Low-error region is highlighted in dashed white line.

The accuracy of this method has been evaluated numerically by starting from a PCF with a known GVD curve calculated with the method detailed in Ref. [15] for the parameters listed in Fig. 1's caption. We numerically calculated the frequency shifts ($\Delta\omega_{S1,2}$) for two pump frequencies ($\omega_{P1,2}$) chosen just below the ZDW, by solving Eq. (1) with the PCF dispersive properties. The ZDW is calculated from Eq.(4) and compared to the original calculated value. Fig.2-(b) shows a color map of the error on the ZDW calculation as a function of the pump power and of the detuning between the pump and the ZDW. This error is less than 1 nm for a pump detuning lower than 15 nm below the ZDW, whatever the pump power up to 100 W (which corresponds to realistic powers). This error increases as the pump wavelength moves away from the ZDW.

### III. EXPERIMENTAL RESULTS

In order to experimentally test the efficiency of our method, we fabricated a PCF with a pitch $\Lambda$ of 4.1 µm and a hole diameter $d$ of 2.6 µm, similar to the parameters used in the previous section. The tunable pump laser was made of a tunable laser diode delivering nanosecond pulses and an Ytterbium-doped fiber amplifier (from Manlight). Spectra were measured after a 10 m-long PCF sample with an optical spectrum analyzer (with a resolution of 0.05 nm) for a pump peak power of 50 W. Figure 3 shows the spectra obtained for two pump wavelengths separated by 0.8 nm ($\lambda_{P1}= 2\pi c/\omega_{P1}=1055.9$ nm and $\lambda_{P2}=2\pi c/\omega_{P2} = 1056.7$ nm). The FWM sidebands are generated around 960 and 1160 nm (these last ones are not shown in this figure for the sake of clarity) with both pump wavelengths.

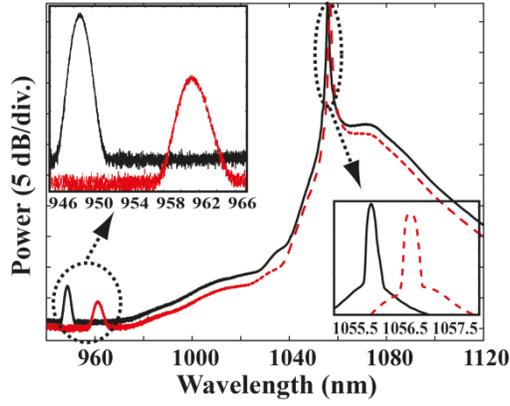

Figure 3: Experimental FWM spectra for pump wavelengths $\lambda_{P1}$ of 1055.9 nm (solid black curve) and $\lambda_{P2}$ of 1056.7 nm (dashed red curve). Left inset: close-up on the signal waves. Right inset: close-up on the pump.

Retrieving the ZDW from Eq. (4) only requires the measurement of the signal wavelength (*i.e.* around 960 nm) for both pump wavelengths. The inset in Fig. 3 shows that the signal is located at 947 nm for $\lambda_{P1}$ and 962 nm for $\lambda_{P2}$, which allows to determine $\Delta\omega_{S1}$ and $\Delta\omega_{S2}$. From Eq. (3) we found that the $\beta_3(\omega_0)/\beta_4(\omega_0)$ ratio is -5.97×10$^2$ ps$^{-1}$ and Eq. (4) gives a ZDW of 1059.4 nm. The reconstruction of the whole GVD curve from Eq. (3) and (4) however requires an additional equation in order to determine independently the third- and fourth-order dispersion coefficients. In the particular case of the pure silica PCF investigated here, we can easily evaluate $\beta_3(\omega_0)$ and $\beta_4(\omega_0)$ with the empirical model of Ref. [15] as follows. Indeed, drawing PCFs unavoidably induces longitudinal fluctuations of chromatic dispersion due to uncontrolled variations of the *d* and $\Lambda$ parameters. Technical specifications of commercial PCFs indeed indicate 10 % fluctuations of the core diameter [16]. Although fluctuations of the microstructured cladding parameters can be significantly reduced to a few % [17], we assume here *d* and $\Lambda$ variations of 10 %. Using the simple model of Ref. [15], we performed 5000 numerical simulations of $\beta_3(\omega_0)$ and $\beta_4(\omega_0)$, by adding a random error with a Gaussian probability to the PCF cladding parameters *d* and $\Lambda$. Results are represented in dots in Fig. 4-(a) and can be well fitted by a third-order polynomial function (green line). Third-and fourth-order dispersion coefficients were found to vary respectively from 6.7 to 7.1×10$^{-2}$ ps$^3$/km and from -1.2 to -0.81×10$^{-4}$ ps$^4$/km at the ZDW. The intersection between the polynomial fit (green line) and the curve corresponding to Eq. (3) (red line) gives $\beta_3(\omega_0)$=6.8×10$^{-2}$ ps$^3$/km and $\beta_4(\omega_0)$=-1.15×10$^{-4}$ ps$^4$/km for our PCF.

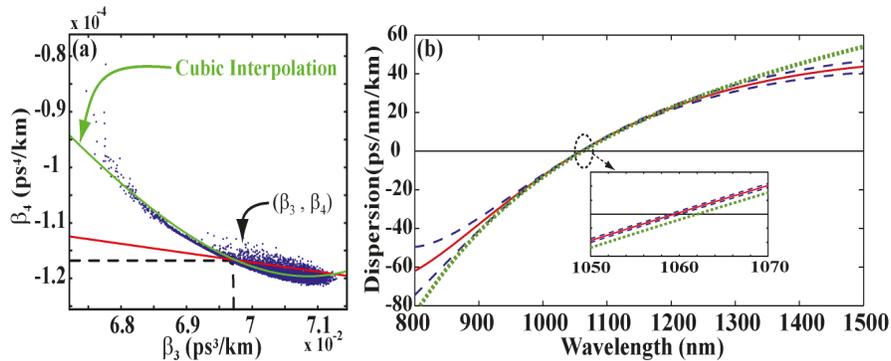

Figure 4: (a) Graphic relationship between $\beta_3$ and $\beta_4$ computed at the ZDW considering a maximum incertitude of 10 % on the PCF cladding parameters ($\Lambda$ and *d*). The red curve corresponds to Eq. (3). (b) GVD curves obtained from our present method (red solid line) and corresponding uncertainty (dashed blue line). The green dotted line corresponds to the GVD curve measured with a low-coherence interferometry setup.

Finally, the GVD curve reconstructed from Eqs. (3), (4) as well as from the results of Fig. 4-(a) is plotted in Fig.4-(b) (red solid line).The estimation of the measurement accuracy was numerically calculated by assuming an uncertainty of 0.1 nm for the measurement of the pump wavelength and of 0.5 nm for the determination of the FWM sidebands. The resulting uncertainty band corresponding to 95% of confidence is plotted in dashed blue lines in Fig.4-(b). The results obtained from our method were then compared with low-coherence interferometry measurements [3] performed on a short sample (1 m) of the same PCF. The resulting GVD curve is plotted in green dotted line in Fig.4-(b) and shows an excellent agreement with the curve reconstructed from our method. In particular, the present method gives a ZDW of 1059.4 nm, against 1062 nm from the low-coherence interferometry measurement. As we can see on this figure, our method is extremely accurate in the low dispersion region and especially for measuring the ZDW of optical fibers. The shortest fiber lengths that can be characterized are typically in the meter range. Indeed, by shortening the fiber, one has to increase the pump power to get the same parametric gain and, as a consequence, the uncertainties on the ZDW become larger (see Fig. 2-(b)). By using longer fibers, one can thus reduce the pump power, but in this case, the growth of the FWM sidebands may rapidly be canceled by longitudinal fluctuations of the fiber dispersion, depending on the fiber quality.

## CONCLUSION

We have presented a simple, accurate and very fast method for measuring the ZDW of optical fibers. We have demonstrated that the measurement of only two spectra for two slightly different pump wavelengths is required to accurately determine the ZDW of an optical fiber as well as the $\beta_3/\beta_4$ ratio. Our method also allows the reconstruction of the GVD curve of a pure silica PCF, in excellent agreement with white-light interferometry measurements from 900 to 1300 nm. This method can be used with fiber lengths ranging from a meter to hundreds of meter. It can be easily used for conventional fibers as long as a tunable source near the ZDW is available, which is the case around 1064 nm or 1550 nm.

## REFERENCES

[1] L. G. Cohen, "Comparison of single mode fiber dispersion measurement techniques", J. Lightwave Technol. **5**, 958-966 (1985).

[2] S. Diddams and J. C. Diels, "Dispersion measurements with white-light interferometry", J. Opt. Soc. Am. B **13**, 1120-1128 (1995).

[3] M. Tateda, N. Shibata and S. Seikai, "Interferometric method for chromatic dispersion measurement in a single-mode optical fibers", IEEE J. Quantum Electron. **17**, 404-407 (1981).

[4] M.J. Saunders and W.B. Gardner, "Interferometric determination of dispersion variations in single-mode fibers", J. Lightwave Technol. **5**, 1701-1705 (1987).

[5] L. F. Mollenauer, P. V. Mamyshev and M. J. Neubelt, "Method for facile and accurate measurement of optical fiber dispersion maps," Opt. Lett. **21**, 1724-1726 (1996).

[6] G. K. Wong et al., "Characterization of chromatic dispersion in photonic crystal fibers using scalar modulation instability," Opt. Express **13**, 8662-8670 (2005).

[7] B. Auguie, A. Mussot, A. Boucon, E. Lantz and T. Sylvestre., "Ultralow chromatic dispersion measurement of optical fibers with a tunable fiber laser," IEEE. Photon. Technol. Lett., **17**, 1825-1827 (2006).

[8] J.M. Chávez Boggio and H.L. Fragnito, "Simple four-wave-mixing-based method for measuring the ratio between the third- and fourth-order dispersion in optical fibers", J. Opt. Soc. Am. B **24**, 2046-2054 (2007).

[9] G. P. Agrawal, Application of Nonlinear Fiber Optics (2003).

[10] S. B. Cavalcanti, J. C. Cressoni, H. R. da Cruz and A. S. Gouvea-Neto, "Modulation instability in the region of minimum group-velocity dispersion of single mode optical fibers via an extended nonlinear Schrödinger equation", Phys. Rev. A, **43**, 6162-6165 (1991).

[11] S. Pitois and G. Millot, "Experimental observation of a new modulation instability spectral window induced by fourth-order dispersion in a normally dispersive single mode optical fiber", Opt. Commun. **226**, 415-422 (2003).

[12] J.D. Harvey et al., "Scalar modulation instability in the normal dispersion regime by use of a photonic crystal fiber", Opt. Lett. **28**, 2225-2227 (2003).

[13] M.E. Marhic, K.K.Y. Wong and L.G. Kazovsky, "Wideband tuning of the gain spectra of one-pump fiber optical parametric amplifiers", IEEE J. Sel. Top. Quantum Electron. **10**, 1133-1141 (2004).

[14] A.Y.H. Chen et al., "Widely tunable optical parametric generation in a photonic crystal fiber", Opt. Lett. **30**, 762-764 (2005).

[15] K. Saitoh and M. Koshiba, "Empirical relations for simple design of photonic crystal fibers," Opt. Express **13**, 267-274 (2005).

[16] NKT Photonics website: http://www.nktphotonics.com/fiber.

[17] B. Stiller et al., "Photonic crystal fiber mapping using Brillouin echoes distributed sensing," Opt. Express **18**, 20136-20142 (2010).